\RequirePackage{fix-cm}
\documentclass[twocolumn,epjc3]{svjour3} 
\smartqed  
\RequirePackage{amssymb}
\RequirePackage{amsmath}
\RequirePackage{array} 
\RequirePackage{epsfig}
\RequirePackage{graphics}
\RequirePackage{colortbl}
\RequirePackage{hhline}
\RequirePackage{multirow}
\begin{document}


\title{Central exclusive diffractive $p$ $\bar{p}$ production in the Regge-eikonal model in the ``scalar'' proton approximation 
}


\author{R.A.~Ryutin \thanksref{e1,addr1}
}

\thankstext{e1}{e-mail: Roman.Rioutine@cern.ch}


\institute{{\small NRC ``Kurchatov Institute'' - Institute for High Energy Physics, Protvino {\it 142 281}, Russia} \label{addr1}
}

\date{
}

\maketitle

\begin{abstract}
Central exclusive diffractive production\linebreak (CEDP) of proton anti-proton pairs was calculated in the Regge-eiko\-nal
approach taking into account continuum and possible $f_0(2100)$ resonance. We use the simple model with the ``scalar'' proton. Data 
from ISR and STAR were analysed and compared with theoretical description. Some predictions for the LHC at 13~TeV are also presented. We discuss shortly possible nuances, problems and pros\-pects of investigations of this process at present and future hadron colliders.
\end{abstract} 
%
%

\section*{Introduction}

As was often mentioned in the papers devoted to the process $p+p\to p+X+p$,
low mass central exclusive diffractive production (LM CEDP) of 
resonances and di-hadron continua
have a lot of advantages to study hadronic diffraction:
\begin{itemize}
\item LM CEDP is the tool for the investigation
of hadronic resonances (like $f_2$ or $f_0$) and their decays to hadrons. We 
can extract different couplings of these resonances to reggeons (pomeron, Odderon etc) to understand their nature (structure and the interaction mechanisms). 
\item We can use LM CEDP to fix the procedure of calculations of ``rescattering'' (unitarity) corrections. For example, in the case of $p$ $\bar{p}$ production we have corrections in the initial proton-proton and final proton-proton and proton-anti-proton channels.
 \item Basic hadrons (pion, proton) are the most fundamental particles in the strong interactions, and LM CEDP gives us a powerful tool to go deep inside their properties, especially to investigate the form factor and scattering amplitudes for the off-shell (``virtual'') hadron.
 \item LM CEDP has rather large cross-sections. It is very important for an exclusive process, since in the special low luminocity runs (of the LHC) we need more time to get enough statistics.
 \item As was proposed in~\cite{myCEDPpipiContinuum},\cite{mySD}, it 
 is possible to extract some reggeon-hadron cross-sections. In the LM \linebreak CEDP of the $p$ $\bar{p}$ we can ana\-ly\-ze properties of the pomeron-pomeron 
  to $p$ $\bar{p}$ exclusive 
  cross-section.
 \item Diffractive patterns of CEDP processes are very sensitive to different
approaches (subamplitudes, form factors, unitarization, reggeization procedure), especially differential cross-sections in $t$ and $\phi_{pp}$ (azi\-mu\-thal angle between final protons), and also $M_{p \bar{p}}$ dependence. That is why these processes are used to verify different models of diffraction~\cite{myCEDP1},\cite{myCEDP2}.
\item Especially in the case of LM CEDP of $p$ $\bar{p}$ we have additional possibilities to investigate spin effects, helicity amplitudes, baryon trajectories, to search for Odderon and extract its coupling to proton. 
\item All the above items are additional advantages provided by the LM CEDP, which has usual properties of CEDP: clear  signature  with  two final protons and two  large  rapidity  gaps  (LRG)~\cite{LRG1},\cite{LRG2}  and the  possibility  to  use  the  “missing  mass  method”~\cite{MMM}.
\end{itemize}

Processes of the LM CEDP of di-hadrons were calculated in some works~\cite{CEDPw1}-\cite{CEDPw6a} which are devoted to most popular models.
Authors have considered phenomenological, nonperturbative, perturbative and mixed approaches in Reggeon-Reggeon collision subprocess.  Nuances of some approa\-ches were analysed in the introduction
 of~\cite{myCEDPpipiContinuum}.

Recently in our works~\cite{myCEDPpipiContinuum},\cite{myCEDPpipiall} we considered the LM CEDP with production of two pions. Here we consider other
possible process, namely, LM CEDP of $p$ $\bar{p}$ system via resonance and 
continuum mechanisms. This process was also considered in~\cite{CEDPw6a} in the tensor pomeron model.

\begin{figure}[hbt!]
\begin{center}
\includegraphics[width=0.49\textwidth]{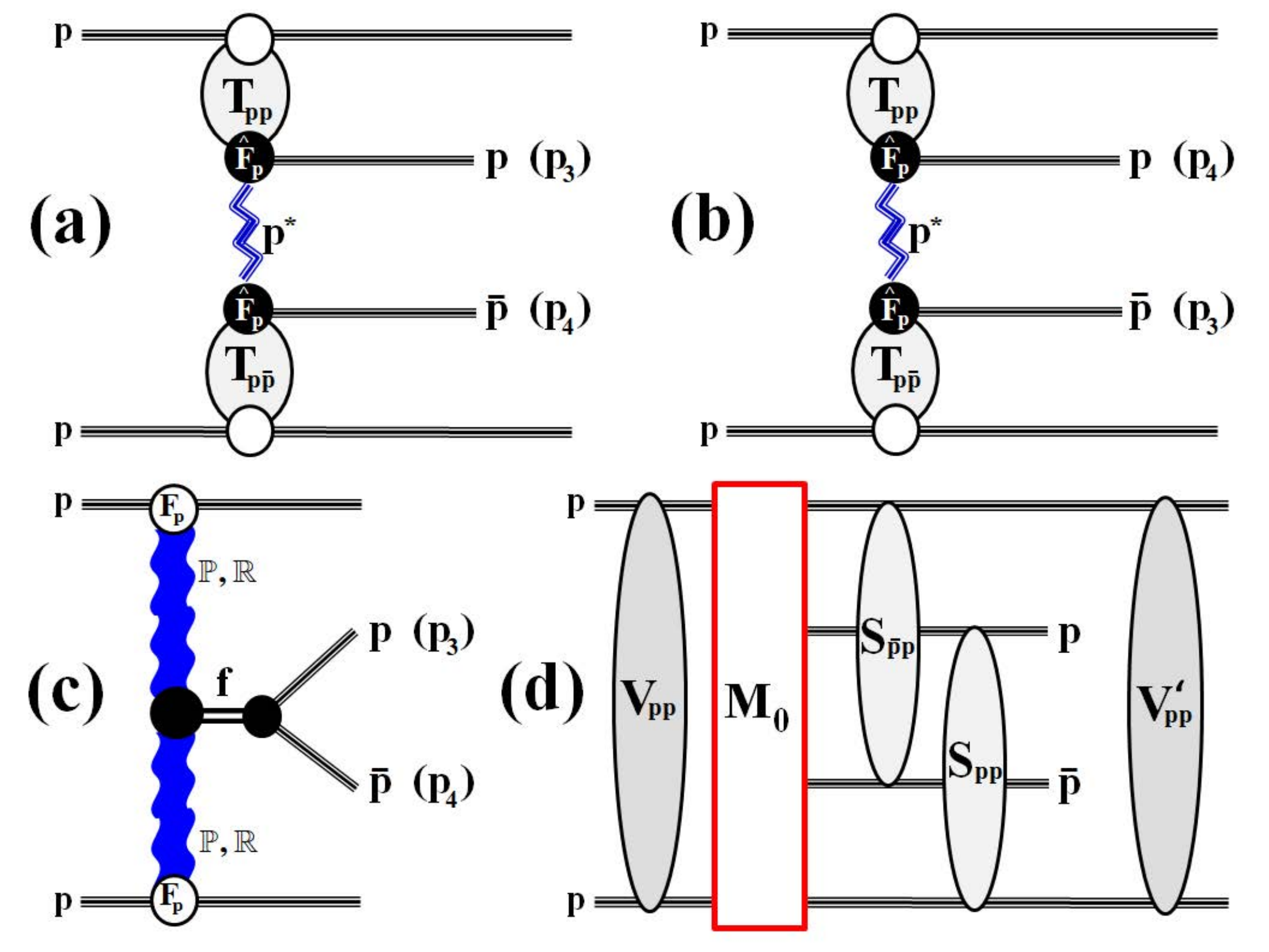}
\caption{\label{fig3:MY4CASES} Amplitudes of the process of LM CEDP $p+p\to p+p\bar{p}+p$ in the Regge-eikonal approach for continuum (a),(b), LM CEDP of $f_0$ resonances (c) with subsequent decay to $p\bar{p}$. (a),(b): central part of the diagram is the continuum CEDP amplitude, where $T_{pp}$, $T_{p\bar{p}}$ are full elastic proton-proton(anti-proton) amplitudes, and the 
proton propagator depicted as dahsed zigzag line.  (c): central part of the diagram contains pomeron-pomeron-resonance fusion with subsequent decay to  $p\bar{p}$, propagator is taken in the Breit-Wigner approximation. Off-shell proton form factor on (a),(b) and other suppression form-factors (in the pomeron-pomeron-f or the $p\bar{p}f_0$) on (c) are presented as a black circles. Full unitarized amplitude (d) contains proton-proton rescatterings in the initial and final states, which are depicted as $V_{pp}$ and $V'_{pp}$-blobes correspondingly, and proton-proton(anti-proton) rescattering corrections, which are also shown as $S_{pp,p\bar{p}}$-blobes. }
\end{center}
\end{figure}

In this article we consider the case, depicted
in Fig.~\ref{fig3:MY4CASES}, and compare the calculations against
the data from \linebreak ISR~\cite{ISRdata1}-\cite{ISRdata3},
STAR~\cite{STARdata3}-\cite{STARdata5}. Also we make
predictions for the LHC.

In the first part of the present work we introduce the framework
for calculations of di-baryon LM CEDP (kinematics, amplitudes, differential
cross-sections) in the Regge-eikonal approach, wich was considered in details in~\cite{myCEDPpipiContinuum},\cite{myCEDPpipiall}. Here we consider
proton (anti-proton) as a ``scalar'', that is why we can not calculate
specific spin effects (like it was done in~\cite{CEDPw6a}). Our 
goal at the present stage is to make preliminary estimations of some general distributions. Extension of the model to particles with any spin will be discussed in futher theoretical works. In this article calculation are almost
similar to our works~\cite{myCEDPpipiContinuum},\cite{myCEDPpipiall} with
some modifications of the amplitudes.

In the second part we analyse the experimental data on the process
at different energies and compare it with our pedictions.

To avoid complicated expressions in the main text, all the basic formulae are placed
to Appendixes.

The purpose of this work is to give some hints and ideas to experimentalists of
what we can expect in the LM CEDP of $p\bar{p}$: possible magnitudes of the couplings and cross-sections (from continuum and resonances), what is the difference between di-pion and other di-hadron production, how big could be spin effects, or contributions of secondary reggeons or Odderon and so on.
  
\section{General framework for calculations of LM CEDP}

LM CEDP is the first exclusive two to four process which is driven basically by the pomeron-pomeron fusion subprocess. It serves as a clear process for 
investigations of resonances like $f_0$, $f_2$ and others with masses less than $5$~GeV. At the noment, for low central masses it is a huge problem to use perturbative approach, that is why we apply the Regge-eikonal method for all the calculations. For proton-proton and proton-anti-proton elastic amplitudes we use the model of~\cite{godizovpp}, \cite{godizovpip}, which describe all the available experimental data on elastic scattering. 

\subsection{Components of the framework}

LM CEDP process can be calculated in the following scheme (see Fig.~\ref{fig3:MY4CASES}):
\begin{enumerate}

\item
We calculate the primary amplitudes of the processes, which are depicted as central parts of diagrams in Fig.~\ref{fig3:MY4CASES}. Here we
consider the case, where the bare off-shell proton propagator in the amplitude for continuum
$p\bar{p}$ production is taken in its simple form (without reggeization)
\begin{equation}
{\cal P}_{p}(\hat{t})=1/(\hat{t}-m_p^2),
\label{eq:ProtonPropagatorR}
\end{equation}
where 
$\hat{t}$ is the square of the momentum transfer between a pomeron and a proton in the pomeron-pomeron fusion process (see Appendix~A for details). 

In the general case, which will be considered in future investigations,
we have to do possible reggeization of the spinor proton with
proton trajectory taken, for example, from~\cite{protontraject1}:
 $$
\alpha_{p}(\hat{t})=-0.4+0.9\hat{t}+0.125\hat{t}^2 .
$$ 
As was noted at the beginning, here
we consider the proton (anti-proton) as a ``spin-0 particle''.

Reggeization of the virtual proton propagator is not 
obvious, since the effect of this is expected to be small and moreover it is not even clear
that we are in the relevant kinematic region ($|\hat{t}|\ll \hat{s}=M_{p\bar{p}}^2 $) to include such corrections for central
production. It was verified also in the calculations presented in this paper. For example, we can use the replacement 
\begin{equation}
\frac{1}{\hat{t}-m_{p}^2}\to \frac{\mathrm{e}^{\alpha_{p}(\hat{t})|\Delta Y|}}{\hat{t}-m_{p}^2},
\label{eq:KMRpionprop}
\end{equation}
as was done in~\cite{CEDPw1}-\cite{CEDPw3}. This expression gives correct ``reggeized'' behaviour in the relevant kinematical region, and 
the usual ``bare'' proton propagator behaviour for small difference between rapidities 
of the final central proton (anti-proton). As to authors of~\cite{CEDPw4}-\cite{CEDPw5a}, we could
use phenomenological expression for vir\-tual proton pro\-pagator like (see~(3.25), (3.26) of~\cite{CEDPw5a} for the pion propagator)
\begin{eqnarray}
&& \frac{1}{\hat{t}-m_{p}^2} F(\Delta Y) + (1-F(\Delta Y)) {\cal P}_{p}(\hat{s},\hat{t}),\nonumber\\
&& F(\Delta Y)=\mathrm{e}^{-c_y \Delta Y},\, \Delta Y=y_{p}-y_{\bar{p}}, 
\label{eq:LIBpionprop}
\end{eqnarray}
to take into accound possible non-Regge behaviour for $\hat{t}\sim\hat{s}/2$, i.e. for
small rapidity separation $\Delta Y$ between final proton and anti-proton. The Regge model 
really does not work in this area or it needs to be modified (as was done, for example, in works~\cite{CEDPw4}-\cite{CEDPw6}, 
with empirical formulae or additional assumptions).

We use also full eikonalized expressions for proton-proton and proton-anti-proton amplitudes, which can be found in the Appendix~B.

\item
After the calculation of the primary LM CEDP amplitudes we have to take into account all possible corrections in proton-proton and proton-anti-proton elastic channels due to the unitarization procedure (so called ``soft survival probability'' or ``rescattering corrections''), which are depicted as $V_{pp}$, $V'_{pp}$ and $S_{pp,p\bar{p}}$ blobes in Fig.~\ref{fig3:MY4CASES}. For proton-proton and proton-anti-proton elastic amplitudes we use the mo\-del of~\cite{godizovpp}, \cite{godizovpip} (see Appendix~B). Possible final interaction between hadrons of the central system is not shown in Fig.~\ref{fig3:MY4CASES}, since we neglect it in the present calculations.

\end{enumerate}

In this article we do not consider so called ``enhanced'' corrections~\cite{CEDPw1}-\cite{CEDPw3}, since they give nonleading contributions in our model due to smallness of the triple pomeron vertex. Also we have no possible absorptive  corrections inside the $p\bar{p}$ central system, since the central mass is low, and also there is a lack of data on this process to define parameters of the model.

Exact kinematics of the two to four process for our case is outlined in Appendix~A.

Here we use the model, presented in Appendix~B for example. You can use another one, which is proved 
to describe well all the available data on proton-proton and proton-anti-proton elastic processes.

\subsection{Continuum $p\bar{p}$ production}

Final expression for the amplitude for the continuum $p\bar{p}$ production 
with initial proton-pro\-ton and final proton-(anti-)proton ``rescattering'' corrections (see Fig.~\ref{fig3:MY4CASES} (a), (b)) can be written as
\begin{eqnarray}
&& M^U\left( \{ p \}\right) = 
\nonumber\\
&&  
=\int\int 
\frac{d^2\vec{q}}{(2\pi)^2}\frac{d^2\vec{q}^{\prime}}{(2\pi)^2}
\frac{d^2\vec{q}_1}{(2\pi)^2}\frac{d^2\vec{q}_2}{(2\pi)^2}
V_{pp}(s,q^2)
V_{pp}(s^{\prime},q^{\prime 2})\nonumber\\
&& \times\; \left[
S_{p\bar{p}}(\tilde{s}_{14},q_1^2)
M_0\left( \{ \tilde{p}\} \right)
S_{pp}(\tilde{s}_{23},q_2^2)+
(3\leftrightarrow 4)
\right] \label{eq:MU}\\
&& M_0\left( \{ p \}\right)=\phantom{I^{2^{2^2}}}\nonumber\\
&&
=T^{el}_{pp}(s_{13},t_1)
{\cal P}_{p}(\hat{t})
\left[ 
\hat{F}_{p}\left( \hat{t}\right)
\right]^2
T^{el}_{p\bar{p}}(s_{24},t_2),
\label{eq:M0}
\end{eqnarray}
where functions are defined in~(\ref{eq:Vpp1})-(\ref{eq:Vpp2}) of Appendix~B, and sets of vectors are
\begin{eqnarray}
&&\{ p \}\equiv \{ p_a,p_b,p_1,p_2,p_3,p_4\}\label{eq:setp}\\
&&\{ \tilde{p}\}\equiv \{ p_a-q,p_b+q; p_1+q^{\prime}+q_1,\nonumber\\
&& \;\;\;\;\;\;\;\;\;\;\;\; p_2-q^{\prime}+q_2,p_3-q_2,p_4-q_1 \} \label{eq:setptild},
\end{eqnarray}
and
\begin{eqnarray}
\tilde{s}_{14}&=&\left( p_1+p_4+q^{\prime} \right)^2,\;
\tilde{s}_{23}=\left( p_2+p_3-q^{\prime} \right)^2,
\label{eq:invarstild}\\
s_{ij}&=&\left( p_i+p_j \right)^2,\;
t_{1,2}=\left( p_{a,b}-p_{1,2} \right)^2,
\label{eq:invars}\\
\hat{s}&=&\left( p_3+p_4 \right)^2,\;
\hat{t}=\left( p_a-p_1-p_3 \right)^2
\label{eq:invarsh}
\end{eqnarray}

Off-shell proton form factor is equal to unity on mass shell $\hat{t}=m_p^2$ and 
taken as exponential
\begin{equation}
\hat{F}_{p}=\mathrm{e}^{(\hat{t}-m_p^2)/\Lambda_p^2},
\label{eq:offshellFpi}
\end{equation}
where $\Lambda_p\sim 1$~GeV is taken from 
the fits to LM CEDP of $p\bar{p}$ at low energies (see next section). In this paper we use only exponential form, but
it is possible to use other parametrizations (see~\cite{CEDPw6a}).

Other functions are defined in Appendix~B. Then we can use the expression~(\ref{eq5:dcsdall}) to calculate
the differential cross-section of the process.

\subsection{CEDP of low mass resonances.}

Here we consider, for example, only one $f_0$ resonance, say, $f_0(2100)$, just
to see the final picture and possible changes in the diffractive patterns.
The general unitarized
amplitude (see Fig.~\ref{fig3:MY4CASES}(c)) is similar to the expression~(\ref{eq:MU}), where amplitude $M_0\left( \{ p \}\right)$ is replaced by the corresponding central primary amplitude for the resonance production and futher decay to $p\bar{p}$. 

 For the $f_0(2100)$ meson amplitude is constructed from the pro\-ton-Pom\-eron form-factor, pomeron-Pome\-ron coupling to the meson\footnote{Here we take the simple scalar one for this meson, although, as was mentioned in our work~\cite{myWA102}, this vertex can be rather complicated and can give nontrivial contribution to the dependence on the azimuthal
 angle between final protons. But for our goals in this paper, namely, investigation of the central mass distributions, it is rather good approximation.}, the off-shell propagator, the off-shell form-factor and the decay vertice. 
 
The primary amplitude is given in the Appendix~C. 
 
\subsection{Nuances of calculations.}

In the next section one can see that there are some difficulties 
in the data fitting. In this 
subsection let us discuss some nuances of calculations, which could change the situation. 

We have to pay special attention to amplitudes, where one or more external particles
are off their mass shell. The example of such an amplitude is the
proton-(anti-)proton one $T_{p p}$ ($T_{p\bar{p}}$), which is the part of the CEDP 
amplitude (see~(\ref{eq:MU})). For this amplitude in the present paper we use Regge-eikonal model with the
eikonal function in the classical Regge form. And ``off-shell'' condition for one of the 
baryons is taken into account by additional phenomenological form factor 
$\hat{F}_{p}(\hat{t})$. But there are at least two other possibilities.

The first one was considered in~\cite{PetrovOffshell}. For amplitude 
with one particle off-shell the formula
\begin{equation}
T^*(s,b)=\frac{\delta^*(s,b)}{\delta(s,b)}T(s,b)=\frac{\delta^*(s,b)}{\delta(s,b)}\frac{\mathrm{e}^{2\mathrm{i}\delta(s,b)}-1}{2\mathrm{i}}
\label{eq:TstarVAP}
\end{equation}
was used. In our case 
\begin{eqnarray}
&& \delta(s,b) = \delta_{p p,p\bar{p}}(s,b; m_p^2,m_p^2,m_p^2,m_p^2),\nonumber\\ 
&& \delta^*(s,b) = \delta_{p p,p\bar{p}}^*(s,b; \hat{t},m_p^2,m_p^2,m_p^2)\nonumber\\
&& \left.\delta_{p p,p\bar{p}}=\delta^*_{p p,p\bar{p}}\right|_{\hat{t}\to m_p^2}.
\label{eq:deltasVAP}
\end{eqnarray}
$\delta_{p p,p\bar{p}}$ is the eikonal function (see~(\ref{eq:elamplitudes})). This is similar to the introduction of the additional form factor, but in a more consistent way, which takes into account the 
unitarity condition.

The second one arises from the covariant reggeization
method, which was considered in the Appendix~C of~\cite{myCEDPpipiContinuum}. For the case of conserved hadronic currents
we have definite structure in the Legendre function, which is transformed in a natural way to the case of the off-shell
amplitude. But in this case off-shell amplitude has a specific behaviour at low t values (see~\cite{myCEDP2} for details). As 
was shown in~\cite{myCEDP2}, unitarity corrections can mask this behavior. Also in
this case we have to take into account the spinor nature of the proton and modify
the covariant reggeization approach presented in~~\cite{myCEDPpipiContinuum}.

\section{Data from hadron colliders versus results of calculations}
\label{sec:data}

Our basic task is to extract the fundamental information on the interaction of hadrons from
different cross-sections (``diffractive patterns''):
\begin{itemize}
\item from t-distributions we can obtain size and shape of the interaction region;
\item the distribution on the azimuthal angle between final protons gives quantum numbers of the
produced system (see~\cite{myCEDP2},\cite{myWA102} and references therein);
\item from $M_c$ (here $M_c=M_{p\bar{p}}$) dependence and its influence on t-dependence we can make some conclusions about
the interaction at different space-time scales and interrelation between them. Also we can extract couplings of reggeons to different resonances.
\end{itemize}

Process $p+p\to p+p\bar{p}+p$  is one of the basic ``standard candles'', which we can use to estimate other CEDP processes. In this section we consider
the available experimental data on the process and make an attempt to extract the information on couplings and form-factors.

\begin{figure}[h!]
\begin{center}
a)\includegraphics[width=0.35\textwidth]{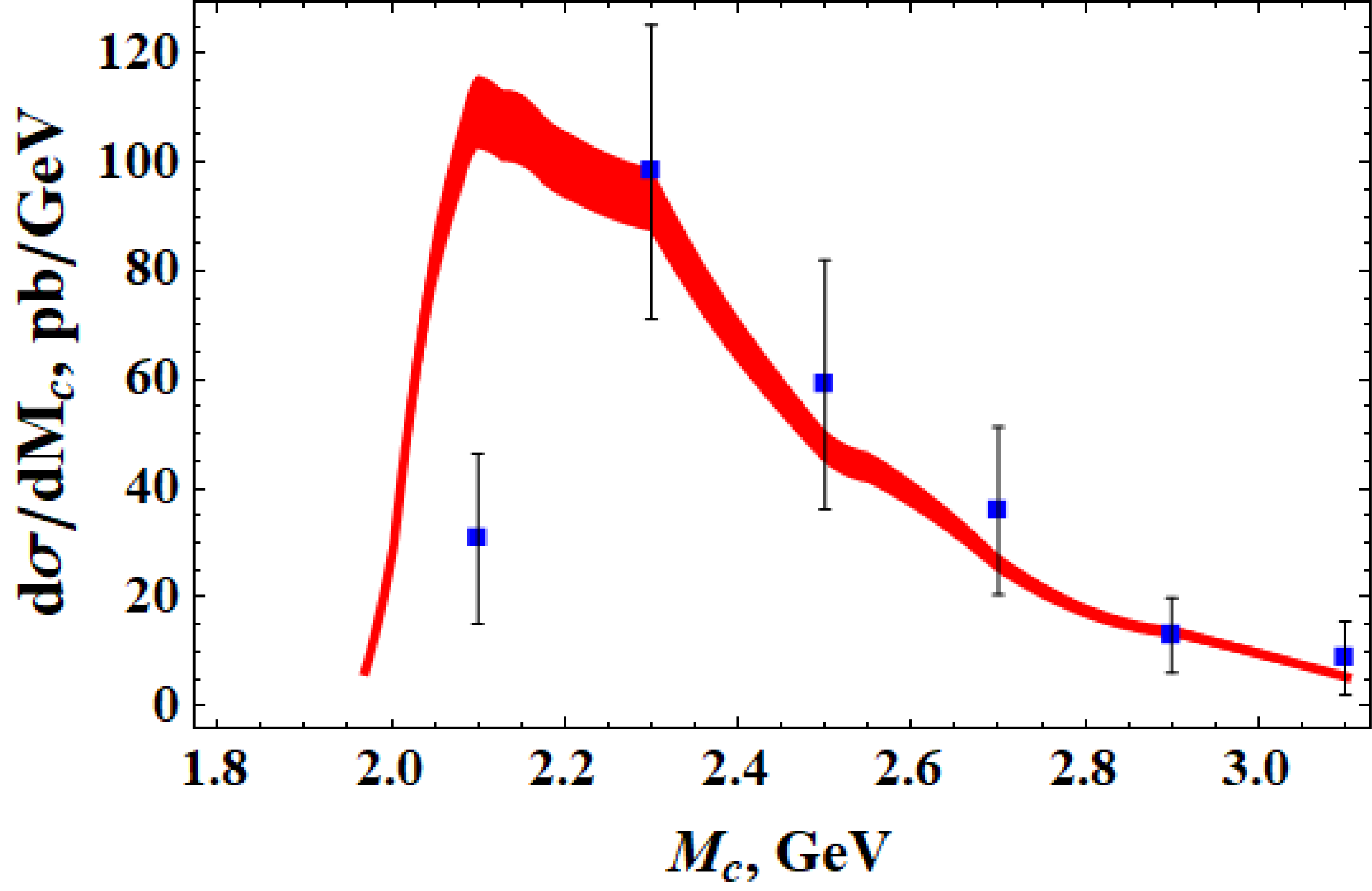}\\
b)\includegraphics[width=0.35\textwidth]{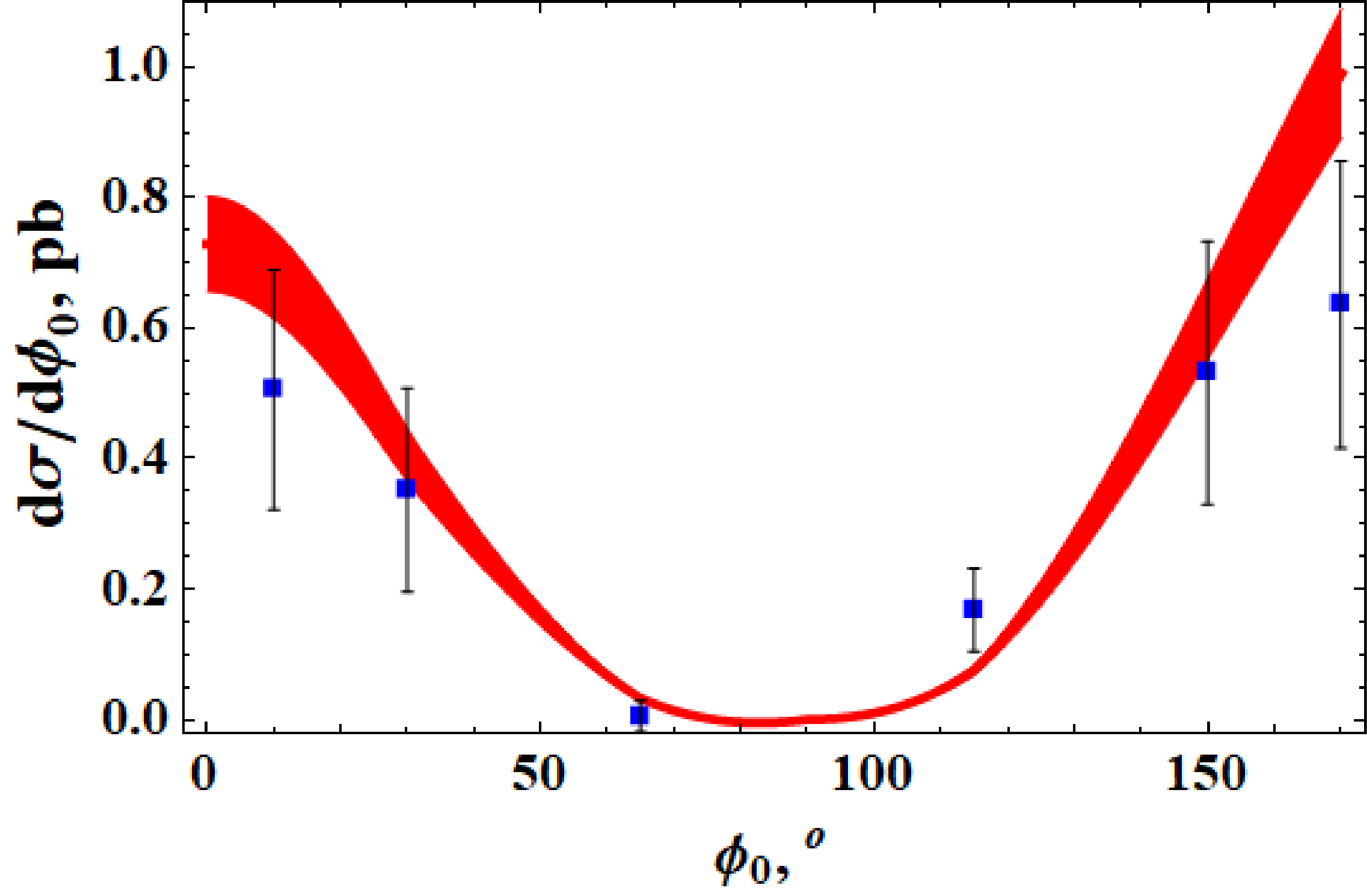}\\
c)\includegraphics[width=0.35\textwidth]{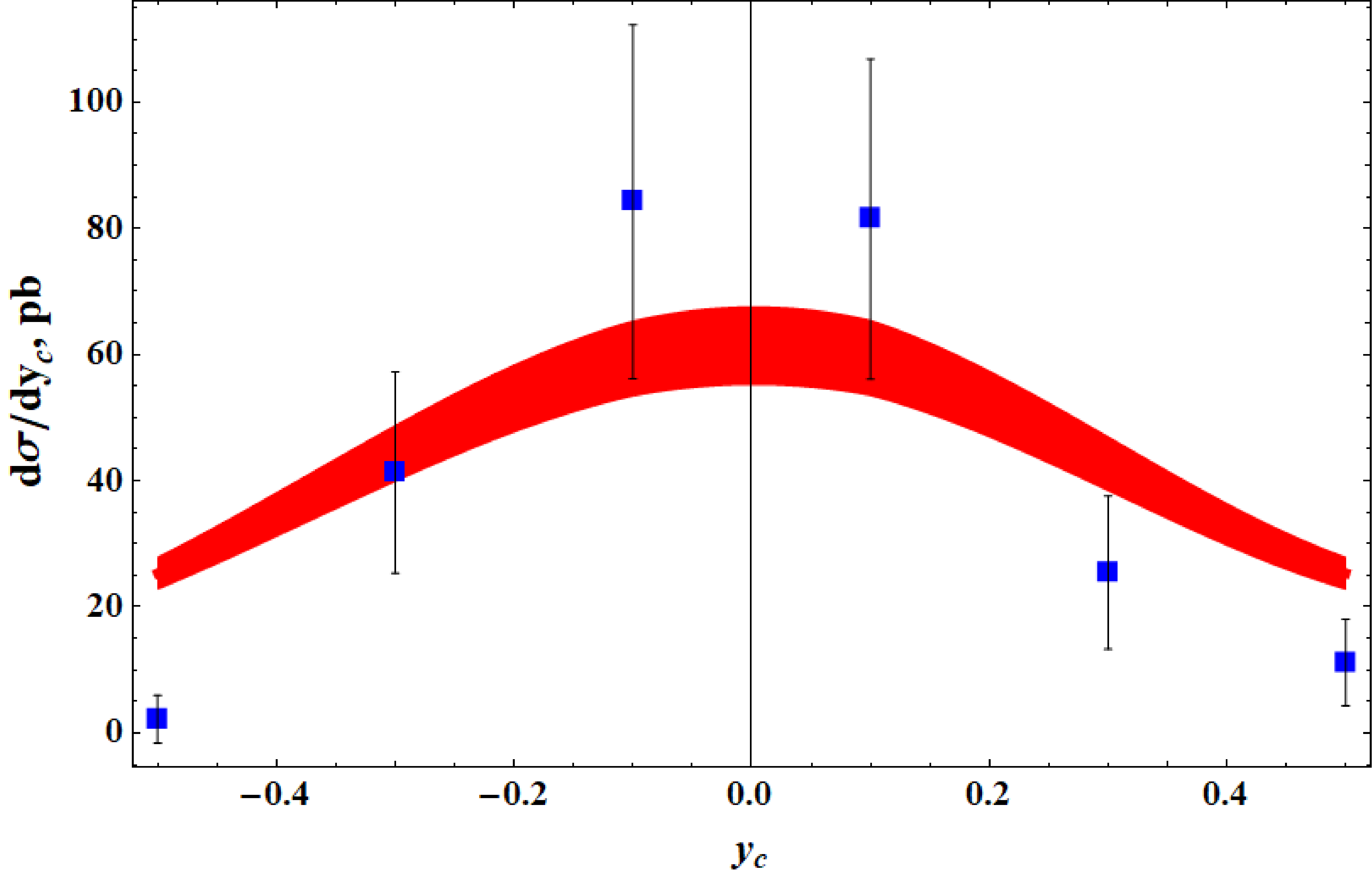}\\
d)\includegraphics[width=0.35\textwidth]{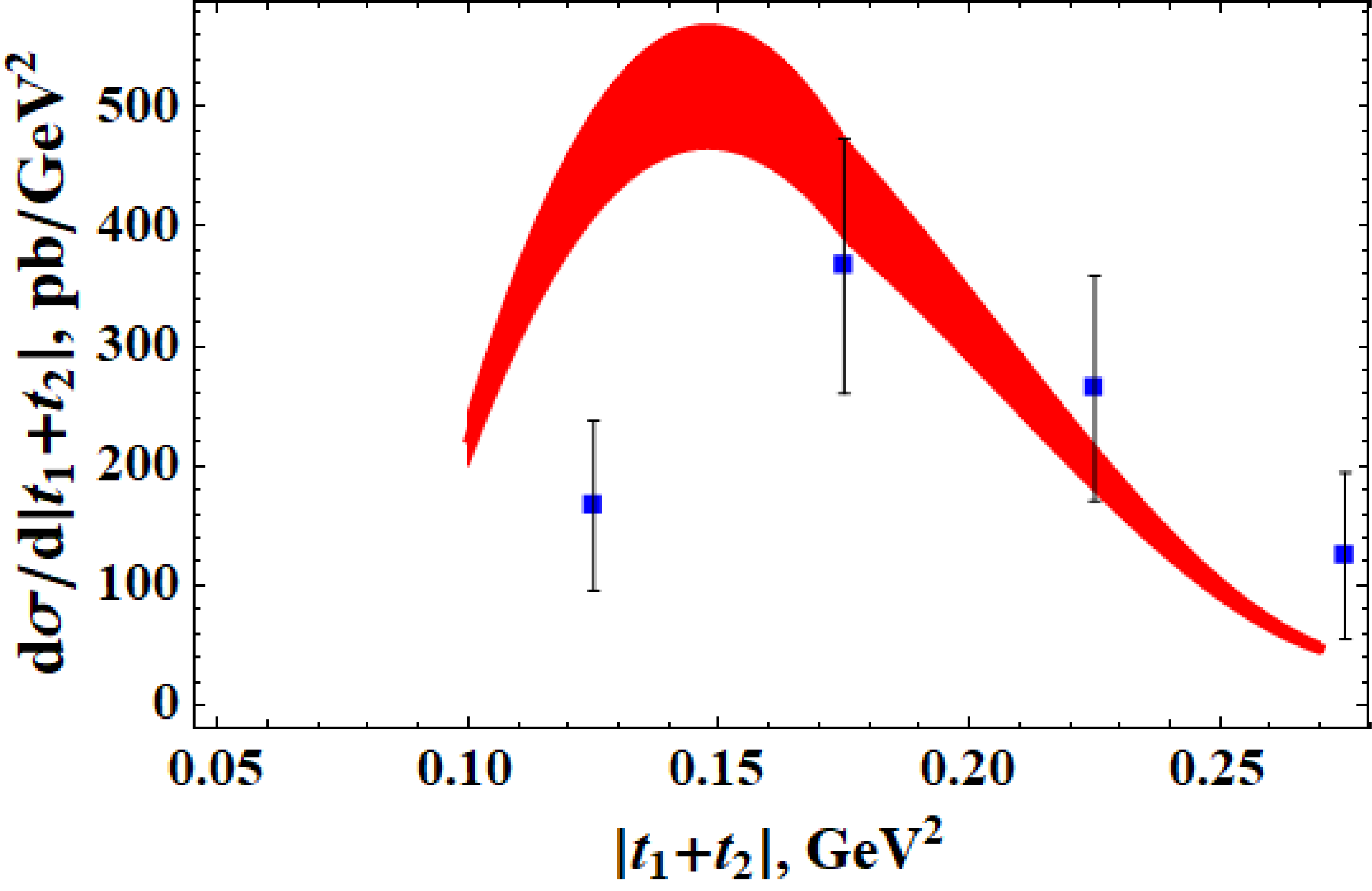}\\
\caption{\label{fig:starNEW} The new data on 
the process $p+p\to p+p\bar{p}+p$ 
at $\sqrt{s}=200$~GeV (STAR collaboration~\cite{STARdata3}-\cite{STARdata5}): $|\eta_{p,\bar{p}}|<0.7$, $p^{p}_{T}(p^{\bar{p}}_T)>0.4$~GeV, $\min(p^{p}_{T},p^{\bar{p}}_T)<1.1$~GeV, $p_x>-0.2$~GeV, $0.2$~GeV$<|p_y|<0.4$~GeV, $(p_x+0.3\,\mathrm{GeV})^2+p_y^2<0.25$~GeV$^2$, where $p$ denotes the momenta of final forward protons.  Curves correspond to $\Lambda_{p}=1.12$~GeV in the off-shell proton form factor~(\ref{eq:offshellFpi}) and pomeron-pomeron-$f_0(2100)$ coupling is equal to $0.64$. Curves correspond to the sum of all amplitudes. Thickness of the curves shows the errors
of numerical Monte-carlo calculations. Additional interpolation was used between calculated points for smoothing.
}
\end{center}
\end{figure}   

\subsection{STAR data versus the model distributions}

In this subsection the data from the STAR collaboration~\cite{STARdata3}-\cite{STARdata5} and model curves for all the cases of Fig.~\ref{fig3:MY4CASES} are presented. In our approach we have two free parameters: $\Lambda_p$ (for the continuum) and the coupling of $f_0(2100)$ to pomeron $g_{\mathbb{PP}f}$, which we can extract from the data or fix from some model assumptions. All the distributions are depicted for 
$\Lambda_p=1.12$~GeV.  Here we take only $f_0(2100)$ resonance with pomeron-pomeron-$f_0(2100)$ coupling is equal to $0.64$ (this value is inspired by the paper on possible ``glueball'' states~\cite{GodizovResonances}).

As you can see from the Fig.~\ref{fig:starNEW}, we can describe the data rather well. From these data we fix the parameter $\Lambda_p = 1.12$~GeV, and then we can make some predictions for other energies. Valuable difference is seen only in the region of small central masses and small $|t_1+t_2|$.

\begin{figure}[h!]
\begin{center}
\includegraphics[width=0.35\textwidth]{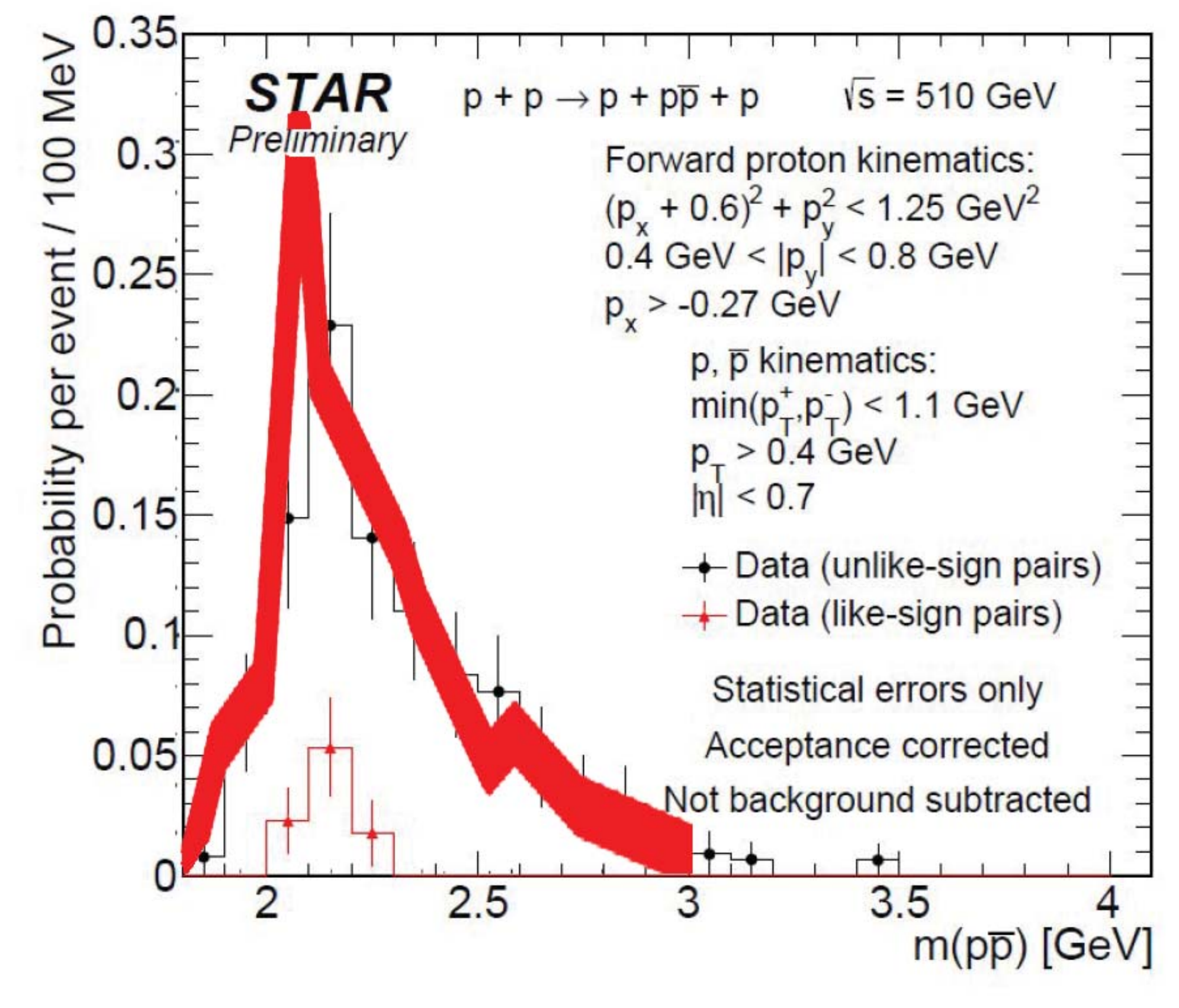}\\
\caption{\label{fig:starNEW510} (a) The new data (normalized cross-section) on 
the process $p+p\to p+p\bar{p}+p$ 
at $\sqrt{s}=510$~GeV (STAR collaboration~\cite{STARdata5}): $|\eta_{p,\bar{p}}|<0.7$, $p^{p}_{T}(p^{\bar{p}}_T)>0.4$~GeV, $\min(p^{p}_{T},p^{\bar{p}}_T)<1.1$~GeV, $p_x>-0.27$~GeV, $0.4$~GeV$<|p_y|<0.8$~GeV, $(p_x+0.6\,\mathrm{GeV})^2+p_y^2<1.25$~GeV$^2$, where $p$ denotes the momenta of final forward protons.  Curves correspond to $\Lambda_{p}=1.12$~GeV in the off-shell proton form factor~(\ref{eq:offshellFpi}) and pomeron-pomeron-$f_0(2100)$ coupling is equal to $0.64$. Thick solid curve corresponds to the sum of all amplitudes, thickness of the curves shows the errors
of numerical calculations.
}
\end{center}
\end{figure}

And on the Fig.~\ref{fig:starNEW510} we see predictions for the STAR data at $\sqrt{s}=510$~GeV.  For the normalized cross-section we see
good description of the distribution shape. Unfortunately, from this data we can not make any conclusions on the absolute value of the cross-section versus
predictions, since for now there are no official publications on the integrated luminosity in this case. So, more correct comparison is postponed for further publications.

\begin{figure}[hbt!]
\begin{center}
\includegraphics[width=0.35\textwidth]{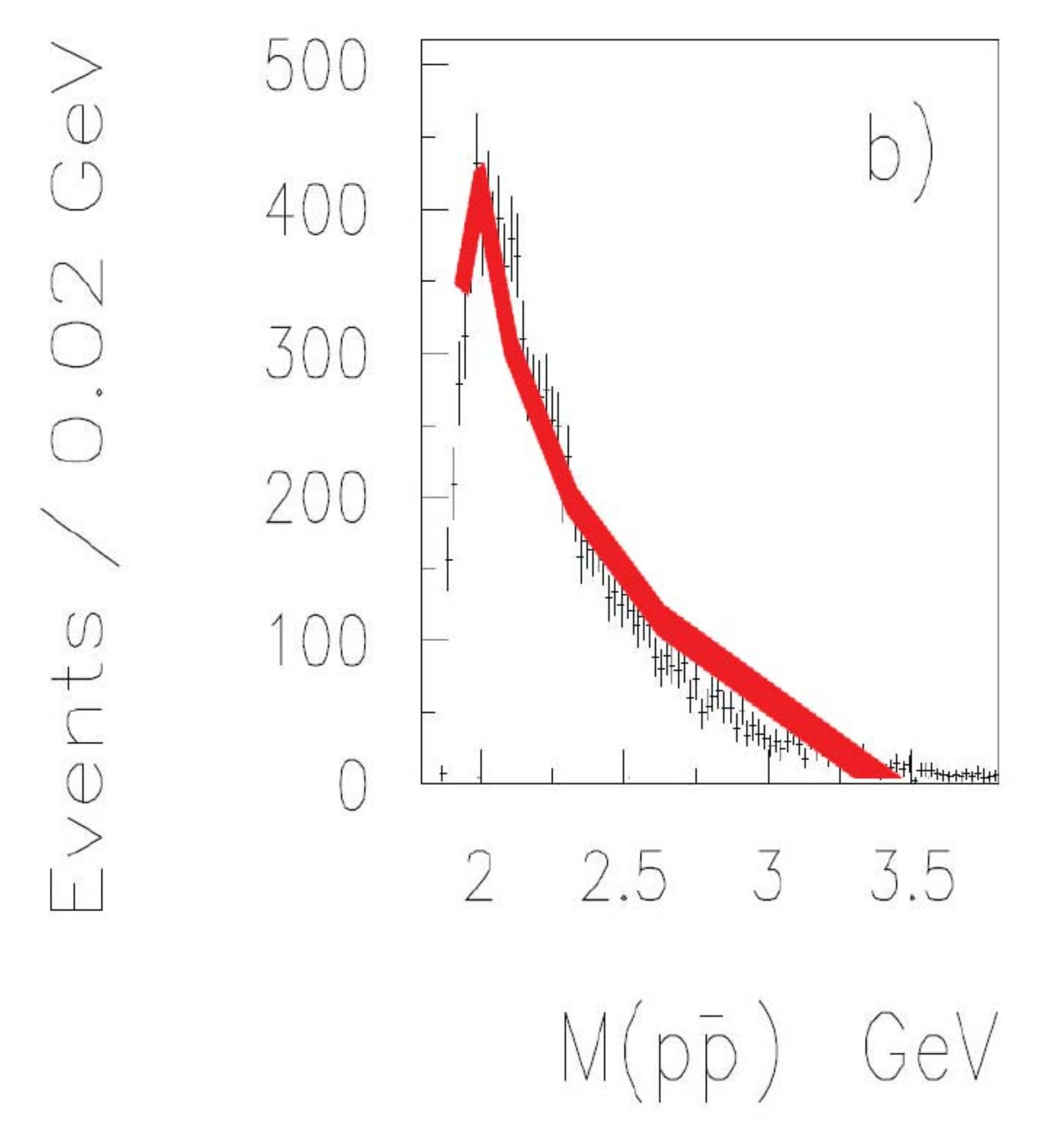}\\
\caption{\label{fig:ISR1exp}  The data on the process $p+p\to p+p\bar{p}+p$  (WA102 collaboration~\cite{ISRdata1}):
at $\sqrt{s}=29.1$~GeV, $|p^c_x|<14$~GeV,$|p^c_y|<0.16$~GeV,$|p^c_z|<0.08$~GeV, $\xi_{1,2\,p}>0.8$.
Theoretical curves, multiplied by 740, correspond to $\Lambda_{p}=1.12$~GeV in the off-shell proton form factor~(\ref{eq:offshellFpi}) and pomeron-pomeron-$f_0(2100)$ coupling is equal to $0.64$. Solid curve corresponds to the sum of all amplitudes, thickness of the curve shows the errors
of numerical calculations.
}
\end{center}
\end{figure}   

\begin{figure}[hbt!]
\begin{center}
\includegraphics[width=0.35\textwidth]{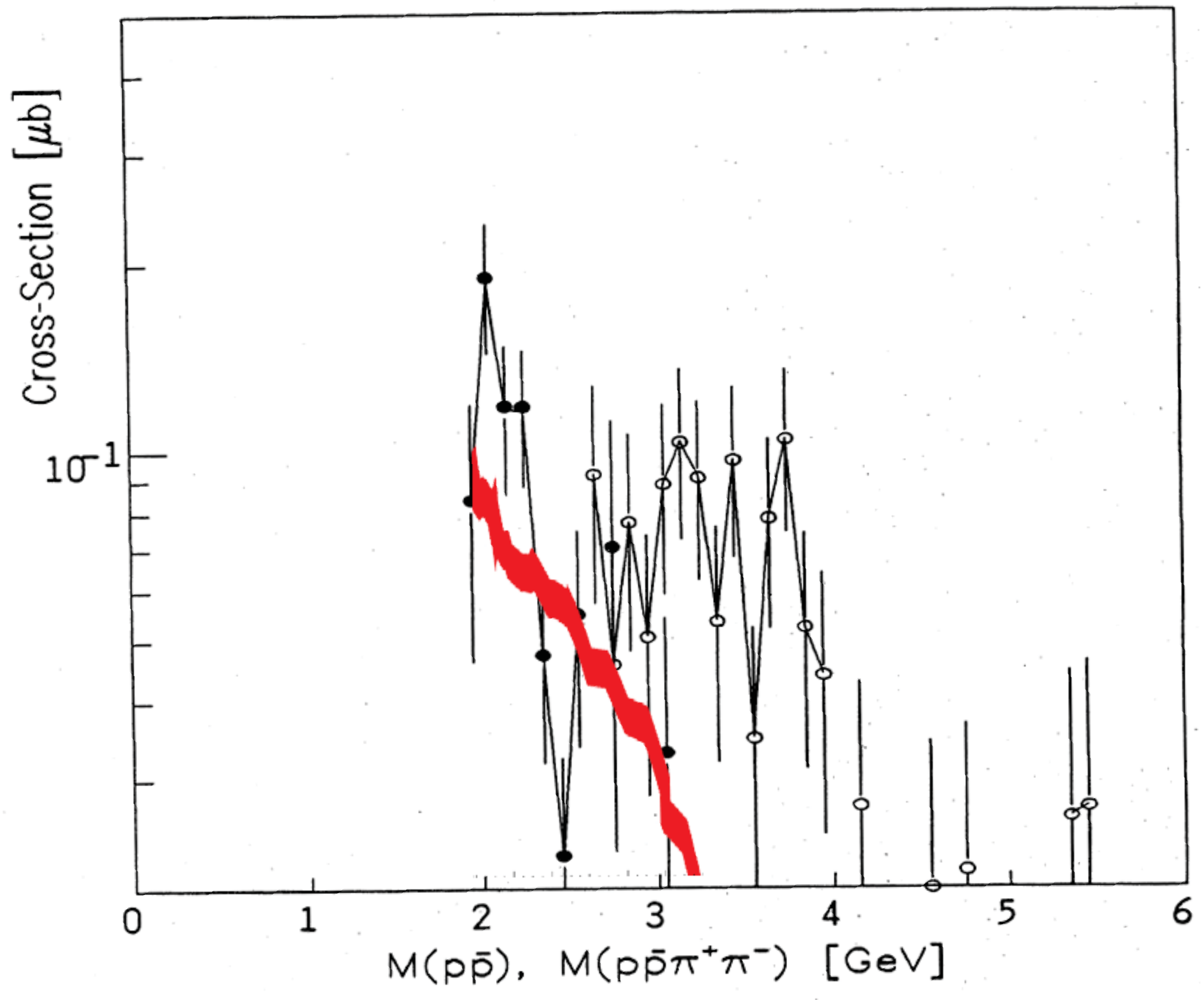}\\
\caption{\label{fig:ISR2exp}  The data (black circles) on the process $p+p\to p+p\bar{p}+p$  (ABCDHW collaboration~\cite{ISRdata2}) at $\sqrt{s}=62$~GeV, $\xi_{1,2\,p}>0.9$, $|y_{p,\bar{p}}|<1.5$, $|t_{1,2}|>0.08$~GeV$^2$. Theoretical curves, multiplied by 200, correspond to $\Lambda_{p}=1.12$~GeV in the off-shell proton form factor~(\ref{eq:offshellFpi}) and pomeron-pomeron-$f_0(2100)$ coupling is equal to $0.64$. Solid curve corresponds to the sum of all amplitudes, thickness of the curve shows the errors of numerical calculations.
}
\end{center}
\end{figure}   

\begin{figure}[hbt!]
\begin{center}
\includegraphics[width=0.35\textwidth]{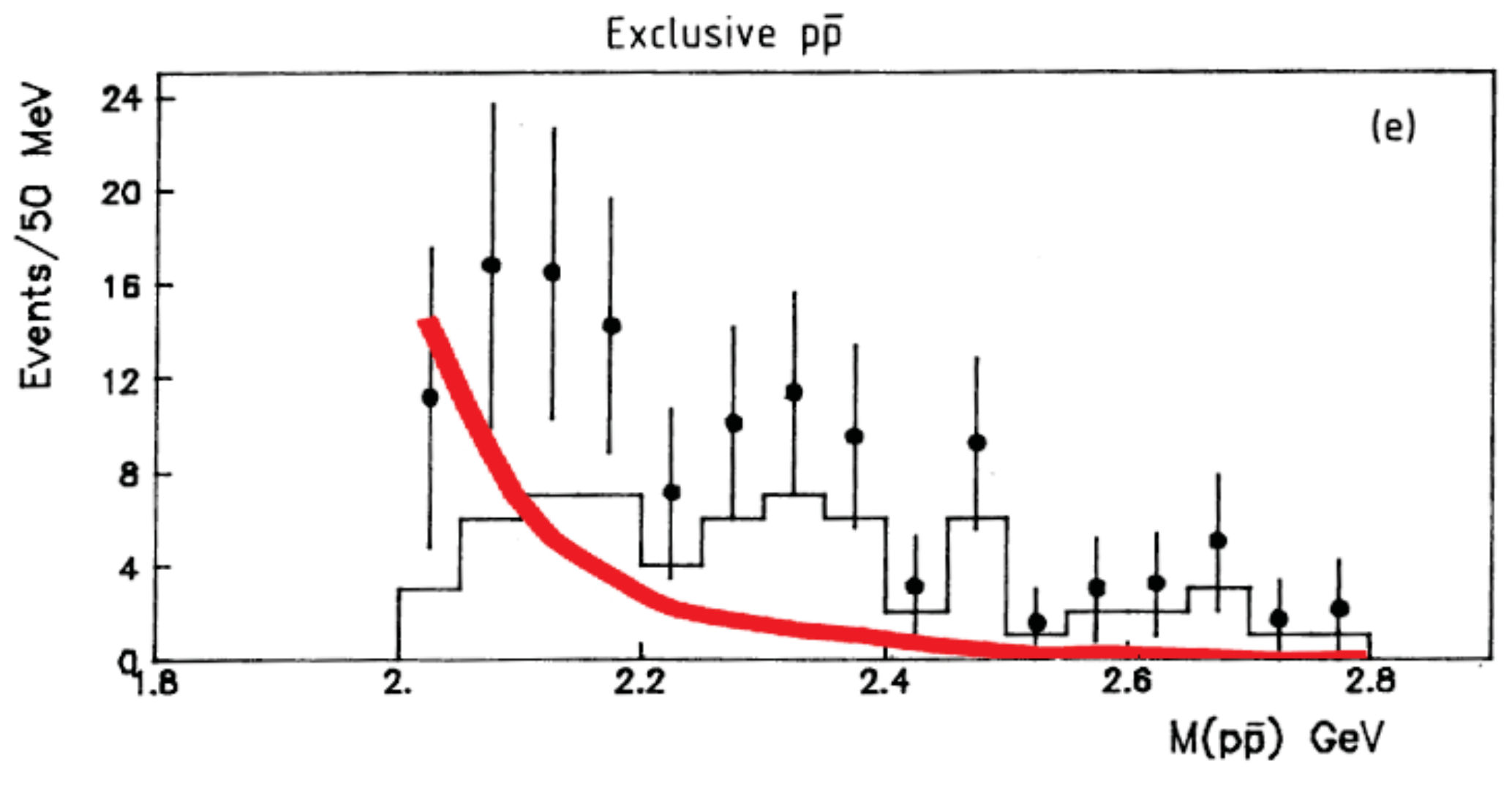}\\
\caption{\label{fig:ISR2newexp}  The data on the process $p+p\to p+p\bar{p}+p$  (AFS collaboration~\cite{ISRdata3}) at $\sqrt{s}=63$~GeV, $\xi_{1,2\,p}>0.95$, $|y_{p,\bar{p}}|<1$, $0.01$~GeV$^2<|t_{1,2}|<0.06$~GeV$^2$.
Theoretical curves correspond to $\Lambda_{p}=1.12$~GeV in the off-shell proton form factor~(\ref{eq:offshellFpi}) and pomeron-pomeron-$f_0(2100)$ coupling is equal to $0.64$.
Solid curve corresponds to the sum of all amplitudes, thickness of the curve shows the errors
of numerical calculations. $\sigma_{exp}=2.5\pm 1.25$~nb was calculated from $d\sigma/dt_1dt_2\sim 1\pm0.5\;\mu$b~GeV$^{-4}$ (see text).
}
\end{center}
\end{figure}   

\subsection{Low energy data versus model cases}

If we take a look at the low energy data, we can 
discover several significant contradictions similar to those found in the production of two pions~\cite{myCEDPpipiall}. This is obvious 
if we use the data of the WA102 collaboration~\cite{ISRdata1} at $\sqrt{s}=29.1$~GeV (see Fig.~\ref{fig:ISR1exp}). The figure shows that the 
shape of the predicted distribution is close to the experimental one, but the \textbf{discrepancy by a factor of 720}
(\textbf{integrated cross-section} is about \textbf{260~pb}, and in~\cite{ISRdata1} we have \textbf{186~nb})
 indicates that the approach used in this energy range fails, and we must take into account other mechanisms for the CEDP of  proton-antiproton pairs. It is possible that resonant production plays a key role here, as was pointed out earlier~\cite{CEDPw6a}, especially 
at low masses of the central system. Also, at low energies, spin effects play a 
significant role, although their contribution will most likely give an increase of no more than $\sim 2$ times.

The situation is almost the same, when we try to compare predictions with the data from ABCDHW collaboration~\cite{ISRdata2} at $\sqrt{s}=62$~GeV (see Fig.~\ref{fig:ISR2exp}). The quality of the data
is not so good, errors are large. \textbf{Experimental 
integrated cross-section} in this case is \textbf{0.8$\pm$0.17~$\mu$b}, and predictions
give only \textbf{0.002}~\textbf{$\mu$b}, i.e. about \textbf{400 times lower}. In this case $|t_{1,2}| > 0.08$~GeV$^2$, and this
cut removes the region which gives the major contribution to the cross-section. We see
the similar discrepancy in~\cite{CEDPw6a}, when one takes into account rescattering corrections.

A more interesting situation arises when we take the data from the AFS collaboration~\cite{ISRdata3}. From this paper we take differential cross sections 
\begin{equation}
\left. d\sigma/dt_1dt_2 \right|_{t_{1,2}=-0.035\;\rm{GeV}^2} 
= 1.0\pm 0.5 
\;\mu\rm{b}\;\rm{GeV}^{-4}.
\label{eq:AFScs}
\end{equation}
If we use a simple form $A\times \rm{e}^{B(t_1+t_2)}$ for the differential cross-section in a very wide range of B
from 3~GeV$^{-2}$ up to 12~GeV$^{-2}$, we can find for the integrated cross-section 
\begin{equation}
\sigma_{exp} \simeq 2.5\pm1.25\;\rm{nb}.
\label{eq:AFScsI}
\end{equation}
 Theoretical calculations give 
\begin{eqnarray}
&& \left. d\sigma/dt_1dt_2 \right|_{t_{1,2}=-0.035\;\rm{GeV}^2} 
= 0.3 \;\mu\rm{b}\;\rm{GeV}^{-4}, \label{eq:thAFScs}\\
&& \sigma_{th}= 1.9\;\rm{nb}.
\label{eq:thAFScsI}
\end{eqnarray}
Also in the Fig.~\ref{fig:ISR2newexp} we can see, that the curve has no such a huge discrepancy with
the experimental points as in two previous cases. The discrepancy is of the same order and even less than
it was in~\cite{myCEDPpipiall} for the di-pion production at the ISR. 

We can conclude, that the model is failed to describe the low energy data, if we fix parameters from the STAR data. For these low energies we have to take into account other mechanisms 
for the CEDP of $p\bar{p}$. Resonant production may play a key role. Also
these mechanisms may include possible corrections 
to proton-proton(anti-proton) amplitudes at low energies, since 
our approach describe data well only for energies greater than $\sim 10$~GeV. And 
in each $T_{pp,p\bar{p}}$ amplitude in Fig.~\ref{fig3:MY4CASES} (a),(b) 
the energy can be even less than $5$~GeV. In the present calculations, just to check preliminary and 
qualitatively the effect of secondary reggeons at very low energy and to 
improve the situation, we use simple exponential parametrization (Born approximation with secondary reggeons) for $T_{pp,p\bar{p}}$ elastic amplitude
to cover the energy value down to the threshold $\sim 2$~GeV.

\subsection{Predictions for LHC}

\begin{figure}[hbt!]
\begin{center}
a)\includegraphics[width=0.35\textwidth]{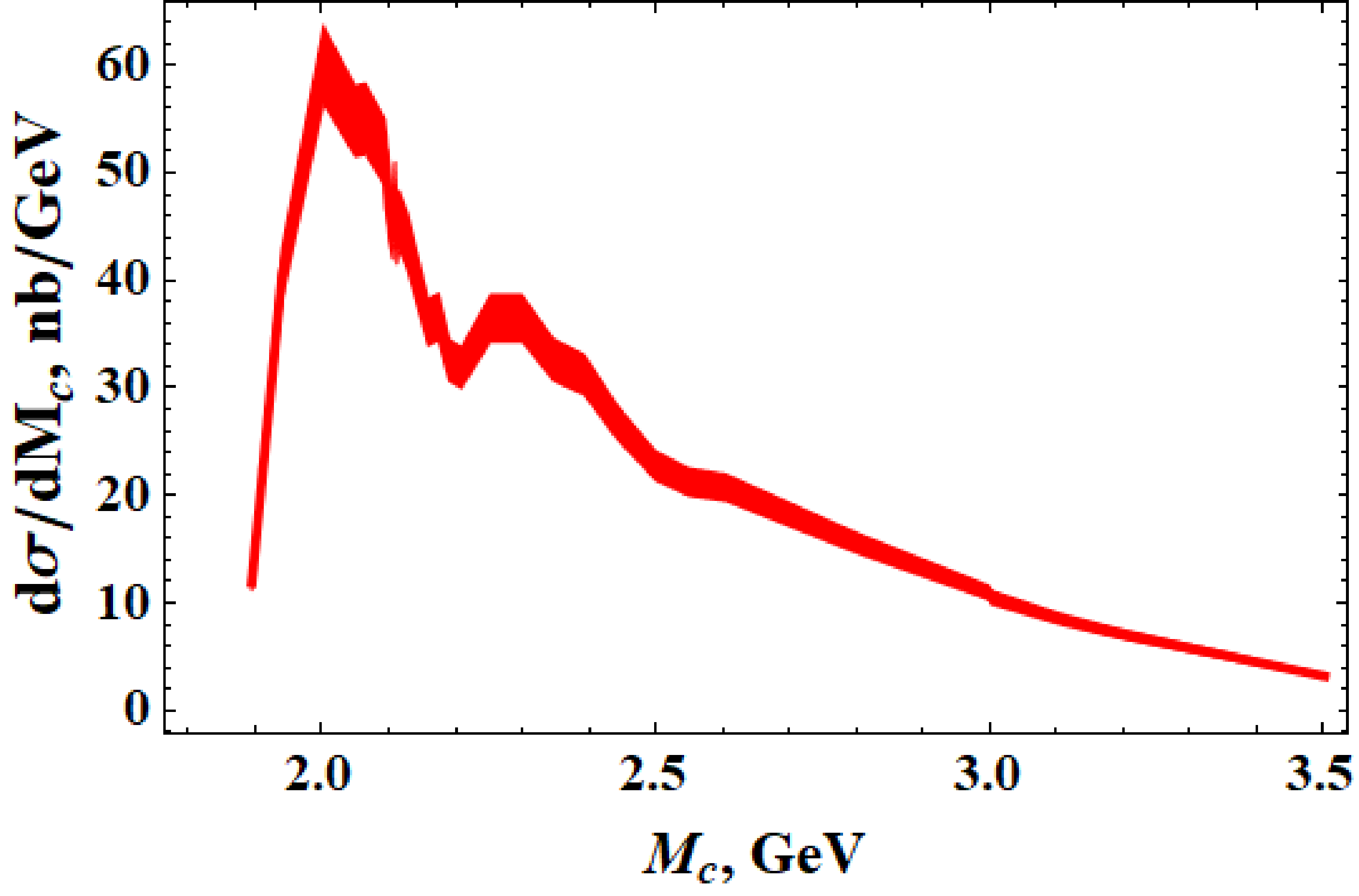}
\caption{\label{fig:CMS} Predictions for CMS energies on the process \linebreak $p+\bar{p}\to p+p\bar{p}+\bar{p}$ are shown 
for $\sqrt{s}=13$~TeV with cuts $|\eta_{p,\bar{p}}|<2.4$, $p_{T,p,\bar{p}}>0.2$~GeV. Theoretical curves correspond to $\Lambda_{p}=1.12$~GeV in the off-shell proton form factor~(\ref{eq:offshellFpi}) and pomeron-pomeron-$f_0(2100)$ coupling is equal to $0.64$.
Solid curve correspond to the sum of all amplitudes, thickness of the curve shows the errors
of numerical calculations.
}
\end{center}
\end{figure}   

In Fig.~\ref{fig:CMS} one can see the prediction for the LHC at 13 TeV for the process, that corresponds to the parameter $\Lambda_p=1.12$~GeV, which 
better fits the STAR data on the Fig.~\ref{fig:starNEW}. The integrated cross-section for these kinematical cuts is about 35~nb, which is about three orders of magnitude smaller than the cross section of di-pion CEDP. The number of $\pi^+\pi^-$ CEDP events is of the order $10^4$ for the LHC at the integrated luminosity
$\sim 200\div 500\;\mu$b$^{-1}$, which is enough to obtain precise distributions in the central mass.  That is why for the CEDP of $p\bar{p}$ we should have at least
$\sim 50\div100$~nb$^{-1}$ integrated luminosity for our investigations.


\section*{Summary and conclusions}

To conclude this article, we can summarize the above analysis in a few statements:
\begin{enumerate}
\item the result is crucially dependent on the choice of $\Lambda_p$ in the off-shell proton form factor, i.e.
on $\hat{t}$ (virtuality of the proton) dependence. This dependance is more significant than in the CEDP of di-pions. In the present approach we take 
$\Lambda_p=1.12$~GeV and couplings:
\begin{eqnarray}
g_{\mathbb{PP}f_0(2100)} &=& 0.64,\nonumber\\
g_{p\bar{p}f_0(2100)} &=& 3.1,
\end{eqnarray}
The coupling of pomeron to $f_0(2100)$ is taken as in~\cite{GodizovResonances} just for tests.

\item if we try to fit the data from STAR~\cite{STARdata3}-\cite{STARdata5}, we can fix the parameter $\Lambda_p=1.12$~GeV, for which the
description is quite good. 

\item for the ISR energies the situation is quite contradictory: 
\begin{itemize}
 \item at $\sqrt{s}=29$~GeV we observe a difference of almost three orders of magnitude (although the shape of the theoretical and experimental distributions 
is the same); 
 \item at $\sqrt{s}=62$~GeV, when 
$|t_{1,2}|>0.08\; \rm{GeV}^2$, 
the difference is already of the order of 200; 
\item at $\sqrt{s}=63$~GeV, for

$0.01\; \rm{GeV}^2<|t_{1,2}|<0.06\;\rm{GeV}^2$,

the predictions turn out to be only about 2 times smaller. 
\end{itemize}

This has to be explained somehow. We can assume
that additional resonant production plays a key role, spin 
effects at low energy are rather big, contributions from other 
processes ($\gamma\gamma\to p\bar{p}$, $\gamma\mathbb{O}\to p\bar{p}$, single and double dissociation) must also be taken 
into account. There are also effects related to the irrelevance and possible modifications of the 
Regge approach (for the virtual proton exchange) in this kinematical region, corrections to
$T_{pp,p\bar{p}}(s,t)$ for\linebreak $\sqrt{s}<5$~GeV, corrections to 
proton-anti-proton scattering at low $M_{p\bar{p}}$;

\item Based on the predictions for $\sqrt{s}=13$~TeV, we can 
say that in order to study this process at the LHC, we need a minimum integrated luminosity of the order of $\sim 50\div100$~nb$^{-1}$.
\end{enumerate}

In further works we will take into account possible modifications of the 
model (proton-anti-proton low energy cross-section, additional 
off-shell effects in subamplitudes, spin effects, contributions 
from dissociative processes and so on) 
for best description of the data. This model will be implemented 
to the new version of the Monte-carlo event generator 
ExDiff~\cite{ExDiffmanual}. It is possible to 
calculate LM CEDP for other di-hadron final 
states ($\phi\phi$, $K^+K^-$, $\eta\eta^{\prime}$ etc.), which 
are also very informative for our understanding of diffractive mechanisms in strong interactions.


\section*{Acknowledgements}

I am grateful to Vladimir Petrov and Anton Godizov for useful discussions and help.  

\newpage
\section*{Appendix A. Kinematics of LM CEDP}
\label{app:kinematics}

\begin{figure}[hbt!]
	\begin{center}
		\includegraphics[width=0.45\textwidth]{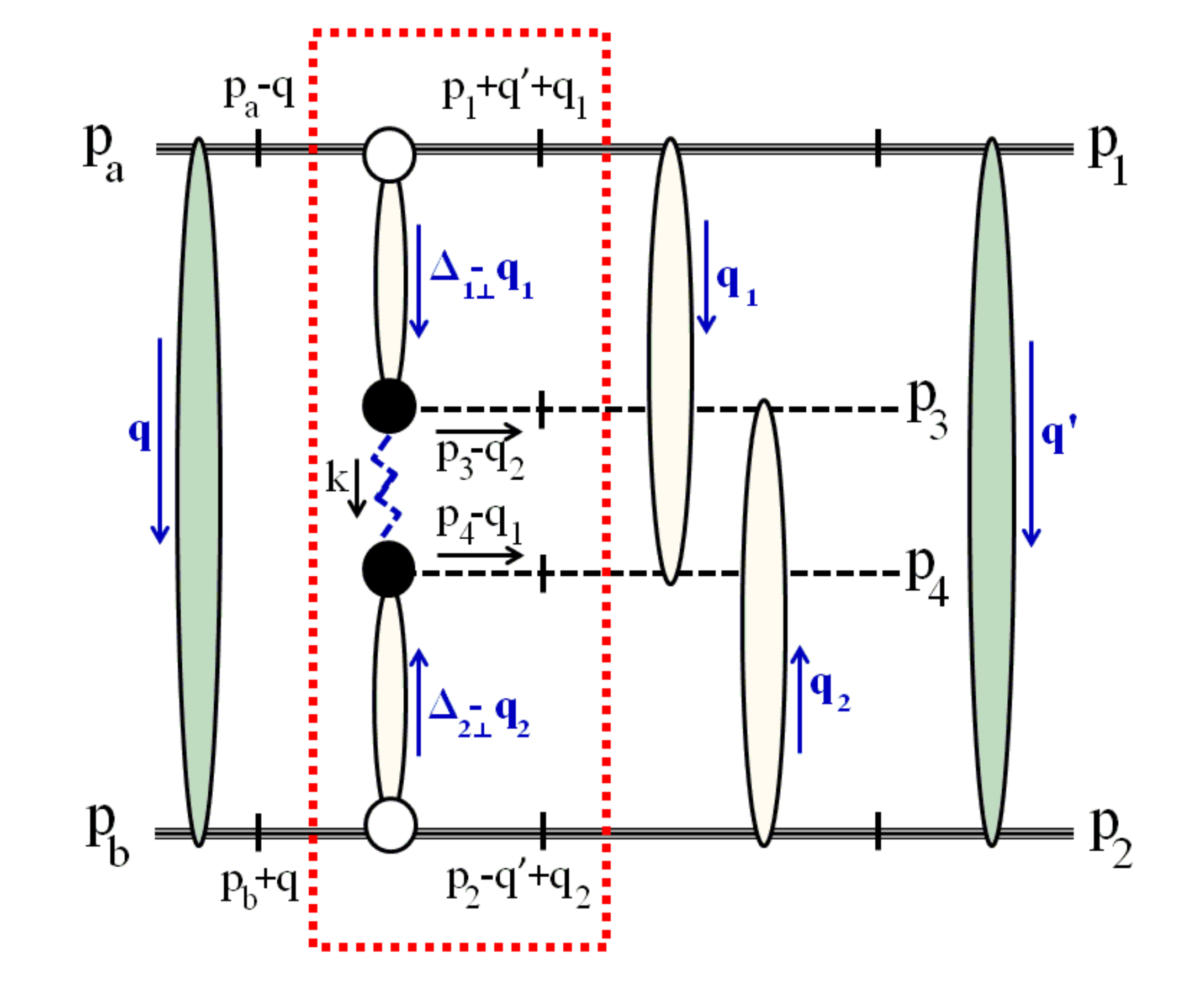}
		\caption{\label{fig4:totalkin} Total amplitude of the process of di-hadron LM CEDP $p+p\to p+h\bar{h}+p$ with detailed kinematics. Proton-proton rescatterings in the initial and final states are depicted as black blobes, and hadron-proton subamplitudes are also shown as shaded blobes. All momenta are shown. Basic part of the amplitude, $M_0$ (see eq.~(\ref{eq:M0})), without corrections is circled by a dotted line. Crossed lines are on mass shell. Here $\Delta_{1\perp}=\Delta_1-q-q^{\prime}$, $\Delta_{2\perp}=\Delta_2+q+q^{\prime}$, $\hat{t}=k^2=(\Delta_{1\perp}-q_1-p_3+q_2)^2$, $\hat{u}=(\Delta_{1\perp}-q_1-p_4)^2$, $\hat{s}=(p_3+p_4-q_1-q_2)^2$.}
	\end{center}
\end{figure}

 The $2\to 4$ process $p(p_a)+p(p_b)\to p(p_1)+p(p_3)+\bar{p}(p_4)+p(p_2)$ can be
described as follows (the notation for any momentum is $k=(k_0,k_z;\vec{k})$, $\vec{k}=(k_x,k_y)$):
\begin{eqnarray}
 p_a&=&\left( 
\frac{\sqrt{s}}{2}, \beta\frac{\sqrt{s}}{2}; \vec{0} 
\right),\;
p_b=\left( 
\frac{\sqrt{s}}{2}, -\beta\frac{\sqrt{s}}{2}; \vec{0} 
\right),\nonumber\\
p_{1,2}&=&
\left( 
E_{1,2},p_{1,2z}; \vec{p}_{1,2\perp}
\right),
E_{1,2}=\sqrt{p_{1,2z}^2+\vec{p}_{1,2\perp}^2+m_p^2},\;\nonumber\\
p_{3,4}&=&
\left( 
m_{3,4\perp}\mathrm{ch}\;y_{3,4}, m_{3,4\perp}\mathrm{sh}\;y_{3,4}; \vec{p}_{3,4\perp}
\right) = \nonumber\\
&=&\left( 
\sqrt{m_{p}^2+\vec{p}_{3,4\perp}^2\mathrm{ch}^2\eta_{3,4}}, |\vec{p}_{3,4\perp}|\;\mathrm{sh}\;\eta_{3,4}; \vec{p}_{3,4\perp}
\right), \nonumber\\
m_{i\perp}^2&=&m_i^2+\vec{p}_{i\perp}^2,\; m_{1,2}=m_p,\; m_{3,4}=m_{h}, \nonumber\\ 
\vec{p}_{4\perp}&=&-\vec{p}_{3\perp}-\vec{p}_{1\perp}-\vec{p}_{2\perp},\;
\nonumber\\
\beta&=&\sqrt{1-\frac{4m_p^2}{s}},\; s=(p_a+p_b)^2,\; s^{\prime}=(p_1+p_2)^2.
\label{eq2:kinmomenta}
\end{eqnarray}
Here $y_i$ ($\eta_i$) are rapidities (pseudorapidities) of final hadrons.

 Phase space of the process in terms of the above variables is the following
\begin{eqnarray}
 d\mathrm{\Phi}_{2\to 4} &=& 
 \left(2\pi\right)^4\delta^4\left( p_a+p_b-\sum_{i=1}^4p_i\right) \prod_{i=1}^4 \frac{d^3p_i}{(2\pi)^32E_i}=
 \nonumber\\
 &=& \frac{1}{2^4(2\pi)^8}
 \prod_{i=1}^3 p_{i\perp}dp_{i\perp}d\phi_i\cdot dy_3dy_4 \cdot {\cal J};
  \nonumber\\
 {\cal J}&=& \frac{dp_{1z}}{E_1} \frac{dp_{2z}}{E_2} 
 \delta\left( \sqrt{s}-\sum_{i=1}^4 E_i\right)
 \delta\left( \sum_{i=1}^4 p_{iz}\right)=\nonumber\\
 &=&\frac{1}{\left| \tilde{E}_2\tilde{p}_{1z}-\tilde{E}_1\tilde{p}_{2z}\right|},
\label{eq3:kinPhSp}  
\end{eqnarray} 
where $p_{i\perp}=\left| \vec{p}_{i}\right|$, $\tilde{p}_{1,2z}$ are appropriate roots of the system
\begin{equation}
\begin{cases}
\label{eq4:sys} &A=\sqrt{s}-E_3-E_4=\sqrt{m_{1\perp}^2+p_{1z}^2}+\sqrt{m_{2\perp}^2+p_{2z}^2},\\
 &B=-p_{3z}-p_{4z}=p_{1z}+p_{2z},
\end{cases}
\end{equation}
\begin{eqnarray}
 \tilde{p}_{1z}&=& \frac{B}{2}+\frac{1}{2(A^2-B^2)}\left[ 
B\left( m_{1\perp}^2-m_{2\perp}^2\right) + A\cdot\lambda_0^{1/2}
\right],\nonumber\\
\label{eq4:sysroot} \lambda_0&=&\lambda\left( A^2-B^2,m_{1\perp}^2,m_{2\perp}^2\right). 
\end{eqnarray}
Here $\lambda(x,y,z)=x^2+y^2+z^2-2xy-2xz-2yz$, and then ${\cal J}^{-1}=\lambda_0^{1/2}/2$.

For the differential cross-section we have
\begin{eqnarray}
\frac{d\sigma_{2\to 4}}{\prod_{i=1}^{3\phantom{I}} dp_{i\perp}d\phi_i\cdot dy_3dy_4 }&=&
\frac{1}{2\beta s}\cdot\frac{\prod_{i=1}^3 p_{i\perp}}{2^4(2\pi)^8\cdot\frac{1}{2}\lambda_0^{1/2}}
\left| T\right|^2=
\nonumber\\
\label{eq5:dcsdall} &=& \frac{\prod_{i=1}^3 p_{i\perp}}{2^{12}\pi^8\beta s \lambda_0^{1/2}}\left| T \right|^2.
\end{eqnarray}
Pseudorapidity is more convenient experimental variable, and we can use the transform
\begin{equation}
\frac{dy_i}{d\eta_i}=\frac{p_{i\perp}\mathrm{ch}\eta_i}{\sqrt{m_i^2+p_{i\perp}^2\mathrm{ch}^2\eta_i}}
\end{equation}
to get the differential cross-section in pseudorapidities.

In some cases it is convenient to use other variables for the integration of the cross-section and
calculation of distributions on central mass. For these cases we have:
\begin{eqnarray}
 d\mathrm{\Phi}_{2\to 4}  &=& \frac{1}{2^4(2\pi)^8}
 \prod_{i=1}^2 dt_i d\phi_i\cdot M_c dM_c d\eta_c dc^* d\phi^*\cdot {\cal J}^{\prime};
  \nonumber\\
 {\cal J}^{\prime}&=& \frac{\beta_M}{4\beta_1\beta_2 s} \frac{dy_c}{d\eta_c},\, \,\,\,\,
 \beta_i \simeq \sqrt{1+\frac{4(m_p^2-(1-\xi_i)t_i)}{\beta^2s^2(1-\xi_i)^2}}; \nonumber\\
\beta_M &=& \sqrt{1-\frac{4m_{p}^2}{M_c^2}},\, \,
\frac{dy_c}{d\eta_c}=\frac{p_{c\perp}\mathrm{ch}\eta_c}{\sqrt{M_c^2+p_{c\perp}^2\mathrm{ch}^2\eta_c}};
\label{eq3:kinPhSpMc}  
\end{eqnarray} 

\begin{eqnarray}
\frac{d\sigma_{2\to 4}}{\prod_{i=1}^{2\phantom{I}} dt_{i}d\phi_i dM_c d\eta_c dc^* d\phi^*}&=&
\frac{1}{2\beta s}\cdot\frac{M_c \beta_M \frac{dy_c}{d\eta_c}}{2^4(2\pi)^8\cdot 4\beta_1\beta_2 s}
\left| T\right|^2
\nonumber\\
\label{eq5:dcsdallMc} &=& \frac{M_c \beta_M \frac{dy_c}{d\eta_c}}{2^{15}\pi^8\beta\beta_1\beta_2 s^2 }\left| T \right|^2,
\end{eqnarray}
where $c^*=\cos\theta^*$, $\theta^*$ and $\phi^*$ are polar and azimuthal angles of the central hadron momenta
in the $h\bar{h}$ rest frame, $M_c$ is the di-hadron mass, $\eta_c$ is the di-hadron 
pseudorapidity, $t_1=(p_a-p_1)^2$, $t_2=(p_b-p_2)^2$ and
$$
\xi_{1,2}\simeq \sqrt{\frac{M_c^2-t_1-t_2+2\sqrt{t_1t_2}\cos(\phi_1-\phi_2)}{s}}
\mbox{\bf e}^{\pm y_c}.
$$

For exact calculations of elastic subprocesses (see Fig.~\ref{fig4:totalkin}) of the type\\ $a(p_1)+b(p_2)\to c(p_1-q_{el})+d(p_2+q_{el})$:
\begin{eqnarray}
q_{el}&=&\left( q_0,q_z; \vec{q}\right),\nonumber\\
q_z&=&-\frac{b}{2a}\left( 1-\sqrt{1-\frac{4ac}{b^2}}\right),\nonumber\\ q_0&=&\frac{A_0q_z+\vec{p}_{1\perp}\vec{q}+\vec{p}_{2\perp}\vec{q}}{A_z},\nonumber\\
a&=&A_z^2-A_0^2,\; b=-2
\left( 
A_z\cdot {\cal D}+A_0 \left( \vec{p}_{1\perp}\vec{q}+\vec{p}_{2\perp}\vec{q}\right)
\right),\nonumber\\
c&=&2 A_z B_z-\left( \vec{p}_{1\perp}\vec{q}+\vec{p}_{2\perp}\vec{q}\right)^2+\vec{q}^2A_z^2,\nonumber\\
A_0&=&p_{1z}+p_{2z},\; A_z=p_{10}+p_{20}, \nonumber\\
B_0&=&p_{1z}\cdot\vec{p}_{2\perp}\vec{q}-p_{2z}\cdot\vec{p}_{1\perp}\vec{q},\nonumber\\
B_z&=&p_{10}\cdot\vec{p}_{2\perp}\vec{q}-p_{20}\cdot\vec{p}_{1\perp}\vec{q},\nonumber\\
{\cal D}&=&p_{1z}p_{20}-p_{2z}p_{10},\label{eq6:qel}
\end{eqnarray}
and $q_{el}^2\simeq -\vec{q}^2$.

\section*{Appendix B. Regge-eikonal model for elastic pro\-ton-pro\-ton and proton-anti-pro\-ton scattering}
\label{app:godizovmodel}

Here is a short review of formulae for the Regge-eikonal approach~\cite{godizovpp}, \cite{godizovpip}, which we use to estimate rescattering corrections in the proton proton and pion proton channels.

Amplitudes of elastic proton-proton and proton-anti-proton scattering are expressed in terms of eikonal functions
\begin{eqnarray}
T_{pp,p\bar{p}}^{el}(s,b)&=&\frac{\mathrm{e}^{-2\Omega_{pp,p\bar{p}}^{el}(s,b)}-1}{2\mathrm{i}},\nonumber\\
\Omega_{pp, p\bar{p}}^{el}(s,b) &=& -\mathrm{i}\,\delta_{pp,p\bar{p}}^{el}(s,b),\nonumber\\
\delta^{el}_{pp,p\bar{p}}(s,b)&=&\frac{1}{16\pi s}\int_0^\infty d(-t) J_0(b\sqrt{-t}) \delta^{el}_{pp,p\bar{p}}(s,t).
\label{eq:elamplitudes},\\
 \delta^{el}_{pp,p\bar{p}}(s,t)&\simeq&\nonumber\\
\!\!\!\!\!\!\!\! g_{pp\mathbb{P}}(t)^2 &&\!\!\!\!\!\!\!\! \left(
\mathrm{i}+\tan\frac{\pi(\alpha_{\mathbb{P}}(t)-1)}{2}) 
\right) \pi \alpha^{\prime}_{\mathbb{P}}(t)
\left( \frac{s}{2s_0}\right)^{\alpha_{\mathbb{P}}(t)},\nonumber\\
 \alpha_{\mathbb{P}}(t)&=&1+\frac{\alpha_{\mathbb{P}}(0)-1}{1-\frac{t}{\tau_a}}\,,\, g_{pp\mathbb{P}}(t)=\frac{g_{pp\mathbb{P}}(0)}{\left( 1-a_g t\right)^2}.
\label{eq:godizovpp}
\end{eqnarray}

Here we use only pomeron contribution for the sake of simpli\-ci\-ty, that is why we can not
analyse possible difference between $pp$ and $p\bar{p}$ elastic scattering due
to contributions of other reggeons.

\begin{table}[ht!]
	\caption{\label{tab1} Parameters for proton-proton(anti-proton) elastic scattering amplitude.}
	\begin{center}
		\begin{tabular}{|c|c|}
			\hline
			Parameter &   Value  \\	
			\hline	
			$\alpha_{\mathbb{P}}(0)-1$ & $0.109$  \\	           	
			\hline
	$\tau_a$ &  $0.535$~GeV$^2$  \\	           	
	\hline	
	$g_{pp\mathbb{P}}(0)$ & $13.8$~GeV  \\	           	
	\hline				
	$a_g$ &    $ 0.23$~GeV$^{-2}$  \\	           	
	\hline					
		\end{tabular}
	\end{center}
\end{table}  

\begin{eqnarray}
V_{pp,p\bar{p}}(s,q^2)&=&\int d^2\vec{b} \,\mathrm{e}^{\mathrm{i}\vec{q}\vec{b}}
\sqrt{1+2\mathrm{i}T_{pp,p\bar{p}}^{el}(s,b)}=\nonumber\\
&=& \int d^2\vec{b}\, \mathrm{e}^{\mathrm{i}\vec{q}\vec{b}} 
\mathrm{e}^{-\Omega_{pp,p\bar{p}}^{el}(s,b)}=\nonumber\\
&=& (2\pi)^2\delta^2\left( \vec{q}\right) + 2\pi \bar{T}_{pp,p\bar{p}}(s,q^2),
\label{eq:Vpp1}\\
\bar{T}_{pp,p\bar{p}}(s,q^2)&=& \int_0^\infty b\,db\, J_0\left( b\sqrt{-q^2}\right)
\left[ 
\mathrm{e}^{-\Omega_{pp,p\bar{p}}^{el}(s,b)}-1
\right]
\nonumber\\
S_{pp,p\bar{p}}(s,q^2)&=&\int d^2\vec{b} \,\mathrm{e}^{\mathrm{i}\vec{q}\vec{b}}
\left(
1+2\mathrm{i}T_{pp,p\bar{p}}^{el}(s,b)
\right)=\nonumber\\
&=& \int d^2\vec{b}\, \mathrm{e}^{\mathrm{i}\vec{q}\vec{b}} 
\mathrm{e}^{-2\Omega_{pp,p\bar{p}}^{el}(s,b)}=\nonumber\\
&=& (2\pi)^2\delta^2\left( \vec{q}\right) + 2\pi \bar{T}_{pp,p\bar{p}}(s,q^2),
\label{eq:Vpp2}
\end{eqnarray}

Functions $\bar{T}_{pp,p\bar{p}}$ are convenient for numerical
calculations, since its oscillations are not so strong.

\section*{Appendix C. Primary amplitude for CEDP $f$ resonance production.}
\label{app:ampres}

Here is the short review of the formulae, which can be obtained by the use of Refs.~\cite{CEDPw6a} and~\cite{GodizovResonances}.

Let us introduce the general diffractive factor
\begin{equation}
F_{\mathbb{P}}(t,\xi) = g_{pp\mathbb{P}}(t)^2\left(
\mathrm{i}+\tan\frac{\pi(\alpha_{\mathbb{P}}(t)-1)}{2}) 
\right) \frac{\pi \alpha^{\prime}_{\mathbb{P}}(t)}{\xi^{\alpha_{\mathbb{P}}}(t)}.
\label{eq:pomerongenfactor}
\end{equation}

For $f_0$ production we have the following expression
\begin{eqnarray}
\phantom{M_{000}}&&\!\!\!\!\!M_0^{pp\to p \{f_0\to p\bar{p} \} p} = \nonumber\\
&=& -F_{\mathbb{P}}(t_1,\xi_1) F_{\mathbb{P}}(t_2,\xi_2) \, g_{\mathbb{P}\mathbb{P}f_0}(t_1,t_2,M_c^2)\times
\nonumber\\
&\times&  
\frac{g_{f_0 p\bar{p}} \left({\cal F}(M_c^2,m_{f_0}^2)\right)^2F_M(t_1)F_M(t_2)}{(M_c^2-m_{f_0}^2)+B_{f_0}(M_c^2,m_{f_0}^2)},
\label{eq:AMPf0GEN}
\end{eqnarray}
where ($M_c > 2m_p$)
\begin{eqnarray}
&& \!\!B_{f_0}(M_c^2,m_{f_0}^2) =\nonumber\\
&&= \mathrm{i}\; \Gamma_{f_0}
\left({\cal F}(M_c^2)\right)^2\left[\frac{1-4m_p^2/M_c^2}{1-4m_p^2/m_{f_0}^2}\right]^{1/2} \label{eq:Bf0BW}\\
&& \!\!\!\! {\cal F}(M_c^2,m_f^2) = F^{\mathbb{P}\mathbb{P}f}(M_c^2,m_f^2)=F^{f  p\bar{p}}(M_c^2,m_f^2)=\nonumber\\
&& \!\!\!\! 
=\exp\left( \frac{-(M_c^2-m_{f}^2)^2}{\Lambda_{f}^4}\right),\, \Lambda_{f}\sim 1\,\mathrm{GeV},
\label{eq:FFf0}\\
&& \!\!\!\! F_M(t) = 1/(1-t/m_0^2),\, m_0^2 = 0.5\,\mathrm{GeV}^2
\label{eq:FMLib}
\end{eqnarray}
${\cal F}(M_c^2,m_f^2)$ and $F_M(t)$ are off-shell phenomenological form-factors introduced in~\cite{CEDPLIBbasis}-\cite{CEDPLIBrho} to make more good description of the data. Here we fix mass and width of $f_0(2100)$ meson as in~\cite{pdgpars}
\begin{eqnarray}
m_{f_0(2100)} &=& 2.086\,\mathrm{GeV},\, \Gamma_{f_0(2100)} = 0.284\,\mathrm{GeV}\label{eq:f02100pars}.
\end{eqnarray}
In this work couplings $g_{f_0 p\bar{p}}$ and $g_{\mathbb{P}\mathbb{P}f_0}$ 
are constants and can be extracted from the experimental data. So, finally we can use the product of these two couplings as free parameter, since both are unknown.
To estimate $g_{\mathbb{P}\mathbb{P}f_0}$ we can use assumptions from~\cite{GodizovResonances}, where it is taken to be $0.64$~GeV for any ``glueball like'' resonance. 

Here we take for all off-shell propagators of resonances simple Breit-Wigner
form, but we can use more complicated expressions, which can be found, for 
example, in~\cite{CEDPLIBbasis}-\cite{CEDPLIBrho}.


\begin{thebibliography}{}
%
%

\bibitem{myCEDPpipiContinuum} R.~Ryutin, {\it Central exclusive diffractive production of two-pion continuum at hadron colliders}, Eur. Phys. J. C. \textbf{79}, 981 (2019).

\bibitem{mySD} V.A.~Petrov, R.A.~Ryutin, {\it Single and double diffractive dissociation and the problem of extraction of the proton–pomeron cross-section}, Int. J. Mod. Phys. A \textbf{31}, 1650049 (2016).

\bibitem{myCEDP1} R.~Ryutin, {\it Exclusive Double Diffractive Events: general framework and prospects}, Eur. Phys. J. C. \textbf{73}, 2443 (2013).

\bibitem{myCEDP2} R.~Ryutin, {\it Visualizations of exclusive central diffraction}, 
Eur. Phys. J. C. \textbf{74}, 3162 (2014).

\bibitem{LRG1}  J.D.~Bjorken, {\it Rapidity gaps and jets as a new-physics signature in very-high-energy  hadron-hadron collisions},  Phys. Rev. D \textbf{47}, 101 (1993). 

\bibitem{LRG2} F.~Abe et al. (CDF Collaboration), {\it Observation of rapidity gaps in $\bar{p}$ $p$ collisions  at 1.8~TeV},  Phys. Rev. Lett. \textbf{74}, 855 (1995).   

\bibitem{MMM} M.G.~Albrow,  A.~Rostovtsev, {\it Searching  for  the  Higgs at hadron colliders using the missing mass method}, FERMILAB-PUB-00-173 (2000), [arXiv: 0009336[hep-ph]].

\bibitem{CEDPw1} L.A.~Harland-Lang, V.A.~Khoze, M.G.~Ryskin, {\it Central exclusive production and the Durham model}, Int. J. Mod. Phys. A \textbf{29}, 1446004 (2014).

\bibitem{CEDPw2} L.A.~Harland-Lang, V.A.~Khoze, M.G.~Ryskin,  W.J.~Stirling, 	
{\it Central exclusive production within the Durham model: a review}, 
Int. J. Mod. Phys. A \textbf{29}, 1430031 (2014).

\bibitem{CEDPw3} L.A.~Harland-Lang, V.A.~Khoze, M.G.~Ryskin, {\it Modeling exclusive meson pair production at hadron colliders}, Eur. Phys. J. C \textbf{74}, 2848 (2014). 

\bibitem{CEDPw4} P.~Lebiedowicz, O.~Nachtmann, A.~Szczurek,	
{\it Tensor pomeron, vector odderon and diffractive production of meson and baryon pairs in proton-proton collisions}, EPJ Web Conf. \textbf{206}, 06005 (2019).

\bibitem{CEDPw5} P.~Lebiedowicz, O.~Nachtmann, A.~Szczurek,	
{\it Exclusive diffractive production of $\pi^+\pi^-$ continuum and resonances within tensor pomeron approach}, EPJ Web Conf. \textbf{130}, 05011 (2016).

\bibitem{CEDPw5a} P.~Lebiedowicz, O.~Nachtmann, A.~Szczurek, 
{\it Central exclusive diffractive production of $K^+K^-K^+K^-$  via the intermediate $\phi\phi$ state in proton-proton collisions}, Phys. Rev. D\textbf{99}, 094034 (2019). 

\bibitem{CEDPLIBbasis} C.~Ewerz, M.~Maniatis, O.~Nachtmann, {\it A Model for Soft High-Energy Scattering: Tensor pomeron and Vector Odderon}, Ann. of Phys.
\textbf{342}, 31 (2014); [arXiv:1309.3478 [hep-ph]].

\bibitem{CEDPLIBf0f2} P.~Lebiedowicz, O.~Nachtmann, A.~Szczurek, {\it Central exclusive diffractive production of $\pi^+\pi^-$ continuum, scalar and tensor resonances in $pp$ and $p\bar{p}$ scattering within tensor pomeron approach}, Phys. Rev. D \textbf{93}, 054015 (2016); [arXiv:1601.04537 [hep-ph]].

\bibitem{CEDPLIBf2} P.~Lebiedowicz, O.~Nachtmann, A.~Szczurek, {\it Extracting the pomeron-pomeron- $f_2(1270)$ coupling in the $pp \to pp\pi^+\pi^-$ reaction through angular distributions of the pions}, Phys. Rev. D \textbf{101}, 034008 (2020);  [arXiv:1901.07788 [hep-ph]].

\bibitem{CEDPLIBrho} P.~Lebiedowicz, O.~Nachtmann and A.~Szczurek, {\it $\rho_0$ and Drell-Soding contributions to central exclusive production of $\pi^+\pi^-$ pairs in proton-proton collisions at high energies}, Phys. Rev. D \textbf{91}, 074023 (2015);  [arXiv:1412.3677 [hep-ph]].

\bibitem{CEDPw6} P.~Lebiedowicz, A.~Szczurek,	
{\it Revised model of absorption corrections for the $pp\to p\pi^+\pi^- p$
process}, Phys. Rev. D \textbf{92}, 054001 (2015).

\bibitem{CEDPw6a} P.~Lebiedowicz, O.~Nachtmann, A.~Szczurek, {\it Central exclusive diffractive production of $p\bar{p}$ pairs in proton-proton collisions at high energies}, Phys. Rev. D \textbf{97}, 094027 (2018); [arXiv:1801.03902 [hep-ph]].


\bibitem{myCEDPpipiall} R.~Ryutin, {\it Central exclusive diffractive production of two-pions from continuum and decays of resonances in the Regge-eikonal model}, [arXiv:2112.13274v2]

\bibitem{ISRdata1}  D.~Barberis et al., (WA102 Collaboration), {\it A study of the centrally produced baryon - anti-baryon systems in pp interactions at 450 GeV/c}, Phys. Lett. B \textbf{446}, 342 (1999); [arXiv:hep-ex/9812022].

\bibitem{ISRdata2} A.~Breakstone et al., (ABCDHW Collaboration), {\it Inclusive pomeron-pomeron interactions at the CERN ISR}, Z. Phys. C \textbf{42}, 387 (1989);  [Erratum: Z. Phys. C \textbf{43} (1989) 522].

\bibitem{ISRdata3} T.~Akesson et al., (AFS Collaboration), {\it A search for glueballs and a study of double pomeron exchange at the CERN Intersecting Storage Rings}, Nucl. Phys. B \textbf{264}, 154 (1986).

\bibitem{STARdata3} J.~Adam et al. (The STAR collaboration), {\it Measurement of the central exclusive production of charged particle pairs in proton-proton collisions at $\sqrt{s} = 200$~GeV with the STAR detector at RHIC}, JHEP \textbf{07}, 178 (2020); [arXiv:2004.11078 [hep-ex]]; \linebreak https://www.hepdata.net/record/ins1792394

\bibitem{STARdata4} W.~Guryn, {\it From Elastic Scattering to Central Exclusive Production: Physics with Forward Protons at RHIC}, Acta Phys. Pol. B \textbf{52}, 217 (2021); [arXiv:2104.15041 [nucl-ex]].

\bibitem{STARdata4a} R.~Sikora (for the STAR Collaboration), {\it Central exclusive production of charged particle pairs in proton-proton collisions at $\sqrt{s}=200$~GeV with the STAR detector at RHIC}, PoS \textbf{ICHEP2020},501 (2021); [arXiv:2011.14400 [hep-ex]]

\bibitem{STARdata5} T.~Truhlar (for the STAR Collaboration), {\it Study of the central exclusive production of $\pi^+\pi^-$, $K^+K^-$ and $p\bar{p}$ pairs in proton-proton collisions at $\sqrt{s}=510$~GeV with the STAR detector at RHIC},  [arXiv:2012.06295 [hep-ex]].

\bibitem{godizovpp} A.A.~Godizov,	
{\it Effective transverse radius of nucleon in high-energy elastic diffractive scattering}, Eur. Phys. J. C \textbf{75}, 224 (2015).

\bibitem{godizovpip} A.A.~Godizov, {\it Asymptotic properties of Regge trajectories and elastic pseudoscalar-meson scattering on nucleons at high energies}, Yad. Fiz. \textbf{71}, 1822 (2008).




\bibitem{protontraject1} M.M.~Brisudova, L.~Burakovsky, T.~Goldman, {\it Effective Functional Form of Regge Trajectories}, Phys. Rev. D \textbf{61}, 054013 (2000); [arXiv:hep-ph/9906293].

\bibitem{PetrovOffshell} V.A.~Petrov, {\it High-energy implications of extended unitarity}, IFVE-95-96, IHEP-95-96, talk given at Blois Conference: 20-24 Jun 1995, Blois, France.

\bibitem{myWA102} V.A.~Petrov, R.A.~Ryutin, A.E.~Sobol and J.-P.~Guillaud,  	
{\it Azimuthal angular distributions in EDDE as spin-parity analyser and glueball filter for LHC}, JHEP \textbf{0506}, 007 (2005).

\bibitem{GodizovResonances} A.A.~Godizov, {\it High-energy central exclusive production of the lightest vacuum resonance related to the soft pomeron}, Phys. Lett. B \textbf{787}, 188 (2018).

\bibitem{ExDiffmanual} R.A.~Ryutin, {\it ExDiff Monte Carlo generator for Exclusive Diffraction. Version 2.0. Physics and manual}, [arXiv:1805.08591 [hep-ph]].

\bibitem{pdgpars} P.A. Zyla et al. (Particle Data Group), Prog. Theor. Exp. Phys. \textbf{2020}, 083C01 (2020).


\end{thebibliography}
\end{document}